\documentclass{PoS}
\newcommand{\beq}{\begin{equation}}
\newcommand{\eeq}{\end{equation}}
\usepackage{graphicx}

\PoS{PoS(LAT2005)133}

\title{Meson correlators in a finite volume near the chiral limit }

\author{
  Hidenori Fukaya\\
  Yukawa Institute for Theoretical Physics,
  Kyoto University,
  Kyoto 606-8052, Japan.
}
\author{
  Shoji Hashimoto\\
  High Energy Accelerator Research Organization (KEK),
  Tsukuba 305-0801, Japan.\\
  School of High Energy Accelerator Science,
  The Graduate University for Advanced Studies (Sokendai),
  Tsukuba 305-0801, Japan.
}
\author{
  \speaker{Kenji Ogawa}\thanks{
KEK-CP-168, YITP-05-53
}\\
  School of High Energy Accelerator Science,
  The Graduate University for Advanced Studies (Sokendai),
  Tsukuba 305-0801, Japan.\\
  E-mail: \email{ogawak@post.kek.jp}
}

\abstract{
  We report on the results of our calculation of meson
  correlators in a finite volume.
  The calculation is carried out in the quenched
  approximation near the chiral limit 
  (down to $m_q = 2.6~\mathrm{MeV}$) using the overlap
  fermion.
  For these small quark masses, the scalar and pseudo-scalar
  correlators are well approximated with a few hundred
  eigenmodes. 
  The results for both connected and disconnected
  correlators are compared with the theoretical predictions
  of quenched chiral perturbation theory.
}

\FullConference{XXIIIrd International Symposium on Lattice Field Theory\\
                 25-30 July 2005\\
                 Trinity College, Dublin, Ireland}
\ShortTitle{Meson correlators in a finite volume near the chiral limit}

\begin{document}

\section{Introduction}
In a finite volume the meson correlators are largely
distorted.
Under an extreme condition such as the $\epsilon$-regime
\cite{Gasser:1987ah}, the correlators do not even show the
familiar exponential fall-off as a function of $t$.
It is, however, still possible to analytically describe the
system using the chiral perturbation theory (ChPT), as long
as the pion mass is small enough.
One-loop calculations in the quenched approximation are
available for both connected and disconnected correlators
\cite{Damgaard:2001js,Damgaard:2002qe}.
Since they are expressed in terms of the usual low energy
constants (LECs) in ChPT, the lattice QCD may be used to
determine their values through the calculation of meson
correlators in the $\epsilon$-regime.

In this work we calculate the quenched meson correlators
near the chiral limit.
We employ the overlap-Dirac operator and calculate low-lying
eigenvalues and eigenvectors in order to investigate the
low-lying mode saturation of the meson correlators.
For the scalar and pseudo-scalar correlators the saturation
is well satisfied, and we can obtain them very close to
the chiral limit without extra computing cost.
We present our result for the pion decay constant $F_\pi$
and chiral condensate $\Sigma$.
A full detail of this work is given in \cite{Fukaya:2005yg}
(see also \cite{Ogawa:2005jn}).
Other works in this direction include
\cite{Bietenholz:2003bj,Giusti:2004yp,Bietenholz:2005ip,Shcheredin:2005ew}.

\section{Simulation methods}
We work on a $10^3 \times 20$ lattice at $\beta=5.85$ in the
quenched approximation.
The physical lattice spacing corresponds to $0.123~\mathrm{fm}$.
We use the overlap-Dirac operator
\begin{equation}
  D_m  =  \left(1-\frac{\bar{a}m}{2}\right)D + m,
  ~~D  =  
  \frac{1}{\bar{a}}\left(1+\gamma_5 \mathrm{sgn} (aH_W)\right),
  ~~H_W=\gamma_5\left(D_W-\frac{1}{\bar a}\right),
  \label{eq:overlap2}
\end{equation}
with $\bar a \equiv a/(1+s)$.
The parameter $s$ is set to 0.6.
In the computation of the sign function (\ref{eq:overlap2}),
we treat 60 lowest eigenmodes of $H_W$ exactly and
approximate the rest by the Chebyshev polynomial (degree
100--200).
The topological sectors are identified unambiguously by
counting the number of zero mode of $D$.
We analyzed 20, 45, 44 and 24 gauge configurations
for topological charge $|Q|$ = 0, 1, 2 and 3,
respectively. 
The 200 lowest eigenvalues and associated eigenvectors of the
overlap-Dirac operator $D$ are calculated using the
implicitly restarted Arnoldi method (implemented in
the ARPACK numerical package). 

\begin{figure}[tbp]
  \begin{center}
    \includegraphics[width=65mm]{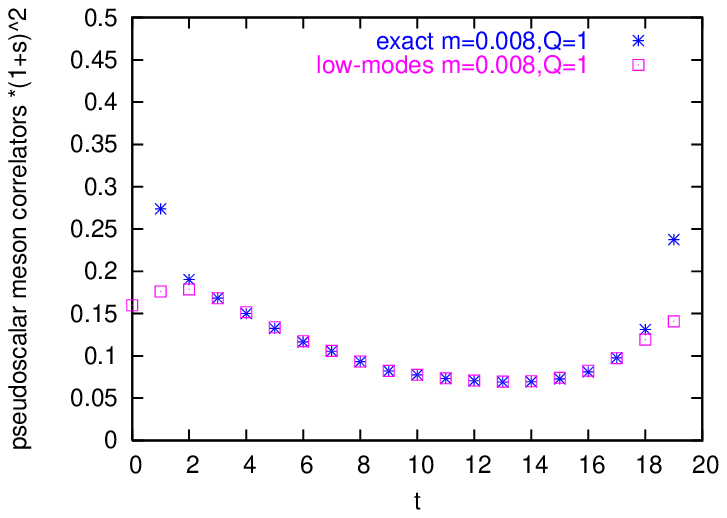}
    \,
    \includegraphics[width=65mm]{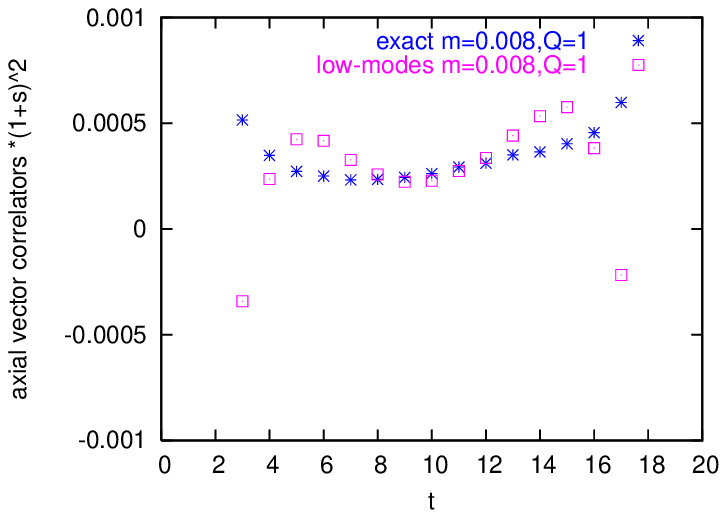}
  \end{center}
  \caption{
    Comparison of the pseudo-scalar (left) and axial-vector
    (right) correlators evaluated with the low-mode
    approximation (stars) with the exactly solved
    correlators (open squares).
  } 
  \label{fig:lowmode}
\end{figure}

Near the massless limit, the quark propagator is expected to
be well approximated by these low-lying eigenvectors. 
\begin{equation}
  D^{-1}_m(x,y)\sim\sum_{i=1}^{200+|Q|}
  \frac{1}{(1-\bar{a}m/2)\lambda_i+m}v_i(x)v_i^{\dagger}(y),
  \label{eq:lowapp}
\end{equation}
where $\lambda_i$ and $v_i(x)$ are the $i$-th eigenvalue and
eigenvector, respectively.
In Figure~\ref{fig:lowmode} we compare the pseudo-scalar
(left panel) and axial-vector (right panel) correlators
constructed from the approximated quark propagator
(\ref{eq:lowapp}) with those from the exact propagator
(obtained using the conventional CG inverter).
At a small quark mass $am$ = 0.008, we find a good agreement
of the approximate correlator with the exact one for the
pseudo-scalar correlator.
Such saturation becomes even better for smaller quark masses
and higher topological sectors.
The saturation is, on the other hand, much worse for the
axial-vector correlator.
Therefore, in the following analysis the axial-vector
correlators are calculated with the exact method, while the
low-mode approximation is used for quantities for which the
saturation is well satisfied.
One advantage of the low-mode approximation
(\ref{eq:lowapp}) is that one can reduce the statistical
noise drastically using the low-mode averaging
technique \cite{Giusti:2004yp,DeGrand:2004qw} without much
extra computing cost.

\section{Axial-vector current correlator}
For the local axial-vector current
$A_\mu^a=\bar\psi\gamma_5\gamma_\mu(\tau^a/2)\psi$ 
we calculate the renormalization factor $Z_A$
non-perturbatively using the Ward-Takahashi identity.
>From the ratio of 
$\langle\bar{\nabla}_0 A^3_0(t) P^3(0)\rangle$
and
$\langle P^3(t)P^3(0)\rangle$
correlators, $Z_A$ is obtained as a linear slope as a
function of $m$.
Our result is $Z_A$ = 1.439(15).

\begin{figure}
  \centering
  \includegraphics[width=65mm]{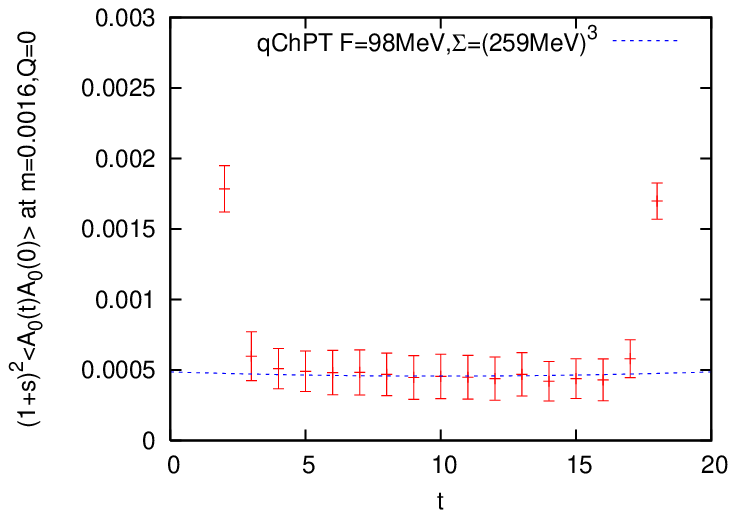}
  \,
  \includegraphics[width=65mm]{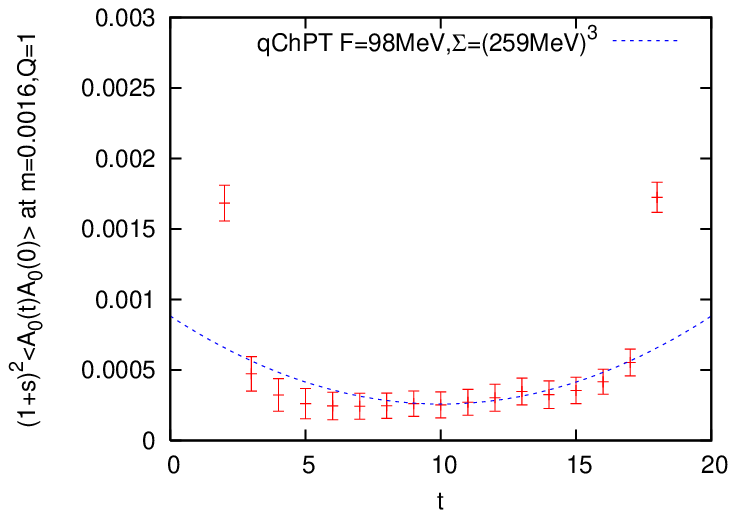}
  \,
  \includegraphics[width=65mm]{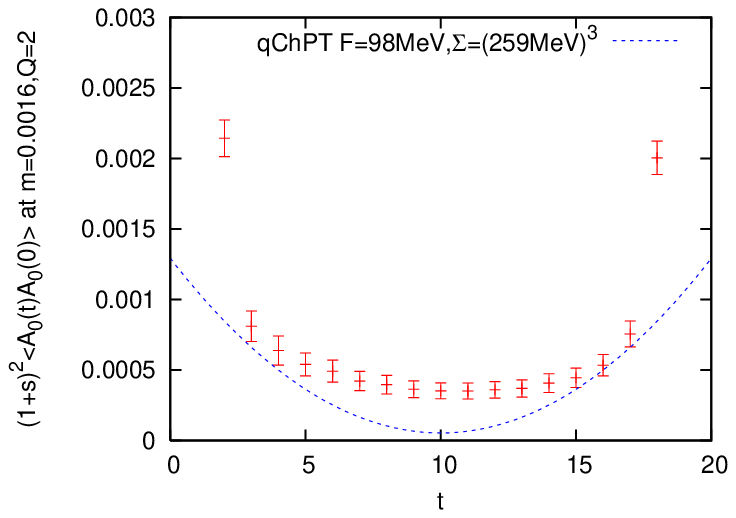}
  \caption{
    Axial-vector current correlator at $Q=0$ (top left),
    $|Q|=1$ (top right) and $|Q|=2$ (bottom).
    The lines show the fit with the function
    (\protect\ref{eq:AA}) for $|Q|$ = 0 and 1.
  }
  \label{fig:axial}
\end{figure}

In Figure~\ref{fig:axial} we show the results for the
axial-vector correlators for topological charges $|Q|$ = 0,
1 and 2.
The expected form from the quenched ChPT (QChPT) is written
as 
\begin{equation}
  \label{eq:AA}
  Z_A^2 \langle A_0^3(t) A_0^{3\dagger}\rangle
  =
  \frac{F_\pi^2}{T} + 2m \Sigma_{|Q|}(\mu) T h_1(t/T).
\end{equation}
where $\Sigma_Q(\mu)$ is the scalar condensate at a given
topological charge $Q$ and $\mu=m\Sigma V$:
$\Sigma_Q(\mu)/\Sigma = \mu
(I_Q(\mu)K_Q(\mu)+I_{Q+1}(\mu)K_{Q-1}(\mu)) + Q/\mu$.
($I_Q(\mu)$ and $K_Q(\mu)$ denote the modified Bessel
functions.) 
The $t$ dependence is simply given by a quadratic function
$h_1(x)=1/2 \left((x-1/2)^2 - 1/12\right)$.
>From this expression, $F_\pi$ is determined through the
constant term, while the second term has dependence on $|Q|$
and $\mu$ as well as $t/T$.
>From a simultaneous fit for the data at $|Q|$ = 0 and 1
with several values of $m$ and $t/T$, we obtain
$F_\pi$ = 98.3(8.3)~MeV and $\Sigma^{1/3}$ = 259(50)~MeV.
The formula (\ref{eq:AA}) explains the data very well at
$|Q|$ = 0 and 1, but the same set of parameters does not
describe the data at higher topological charge ($|Q|$ = 2,
shown in the right panel of Figure~\ref{fig:axial}).
This is not inconsistent with the theoretical expectation,
because the derivation of (\ref{eq:AA}) assumes a condition
$|Q|\ll \langle Q^2\rangle$ and for our small lattice
$\langle Q^2\rangle$ = 4.34(22).
For more detail, see \cite{Fukaya:2005yg}.

\section{Connected scalar and pseudo-scalar correlators}
\begin{figure}
  \begin{center}
    \includegraphics[width=65mm]{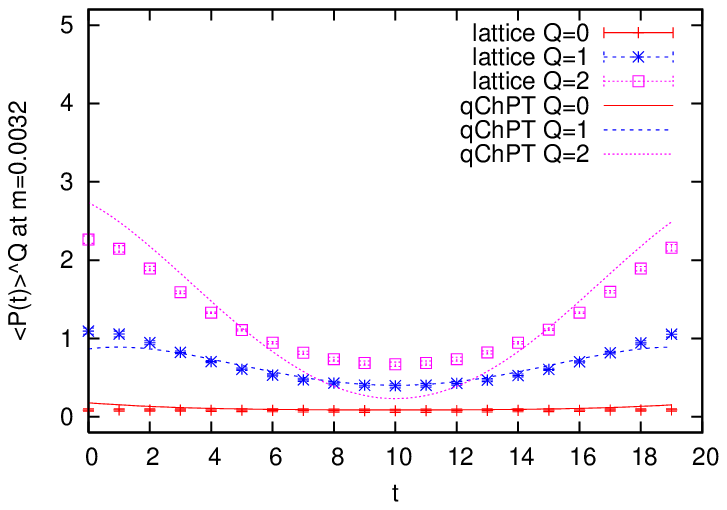}
    \,
    \includegraphics[width=65mm]{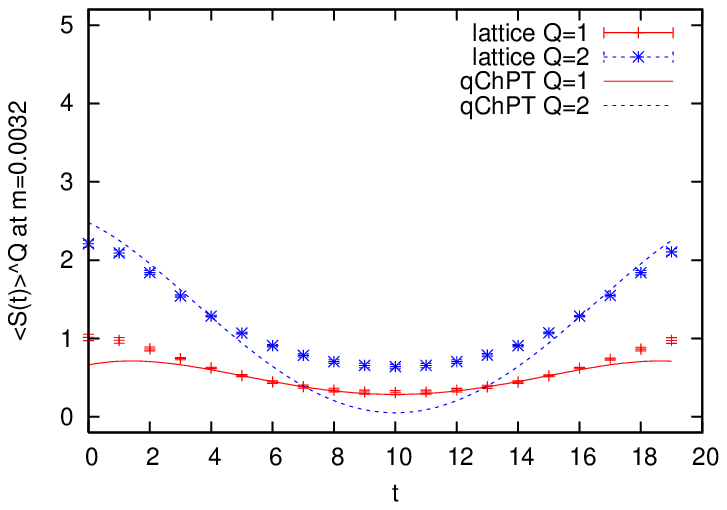}
  \end{center}
  \caption{
    Connected pseudo-scalar (left) and scalar (right
    ) correlators.
    The curves show the fitting with the analytically
    predicted form. 
  } 
  \label{fig:con}
\end{figure}

The connected scalar and pseudo-scalar correlators are
obtained using the low-mode approximation.
Since we can average the source point over the space-time
lattice, the signal is very clean as shown in
Figure~\ref{fig:con}.

The fitting with the QChPT expectation is more
involved than the case of the axial-vector current
correlator, because the parameters $m_0^2$ and $\alpha$
appear to describe the quenched artifacts.
To avoid a four-parameter fit with ($F_\pi$, $\Sigma$,
$m_0^2$, $\alpha$), we input $F_\pi$ from the result of the
axial-vector current correlator and use the relation
$\langle Q^2\rangle/V=m_0^2 F_\pi/2N_c$
to determine $m_0^2$.

The data shown in Figure~\ref{fig:con} are well described by 
the QChPT expectation for $|Q|$ = 0 and 1, as in the
axial-vector current correlator.
We obtain $\Sigma^{1/3}$ = 257(14)(00)~MeV 
and $\alpha$ = $-$4.5(1.2)(0.2), where the first error is
statistical while the second represents the uncertainty of
input parameters.  

Our result for $\alpha$ is rather large and negative, which
contradict with the small $\alpha$ observed in many previous
lattice calculations (for example, see \cite{Bardeen:2003qz}). 
We also attempted a fit with $F_\pi$ and $\Sigma$ as free
parameters, while $\alpha$ is set equal to zero.
The result is $F_\pi$ = 136.9(5.3)(0.9)~MeV 
and $\Sigma^{1/3}$ = 250(13)(00)~MeV;
$F_\pi$ is not consistent with the analysis of the
axial-vector current correlators.

\section{Disconnected pseudo-scalar correlator}
\begin{figure}
  \centering
  \includegraphics[width=65mm]{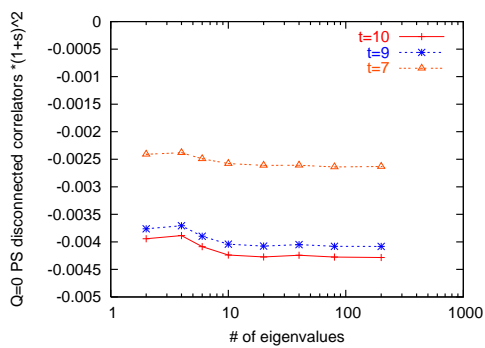}
  \caption{
    Saturation of the disconnected correlator at $t$ = 7, 9 
    and 10 as a function of the number of eigenvalues in the
    lowmode approximation.
  }
  \label{fig:dissat}
\end{figure}

We also study the disconnected pseudo-scalar correlator
using the low-mode approximation.
For this quantity we did not check the consistency with the
exact solution because it is too costly.
Instead, we looked at the saturation as the number of
eigenmodes included in the approximation.
Figure~\ref{fig:dissat} shows that the disconnected diagram
is well saturated with the number of eigenvalues greater
than 10.

\begin{figure}
  \centering
  \includegraphics[width=65mm]{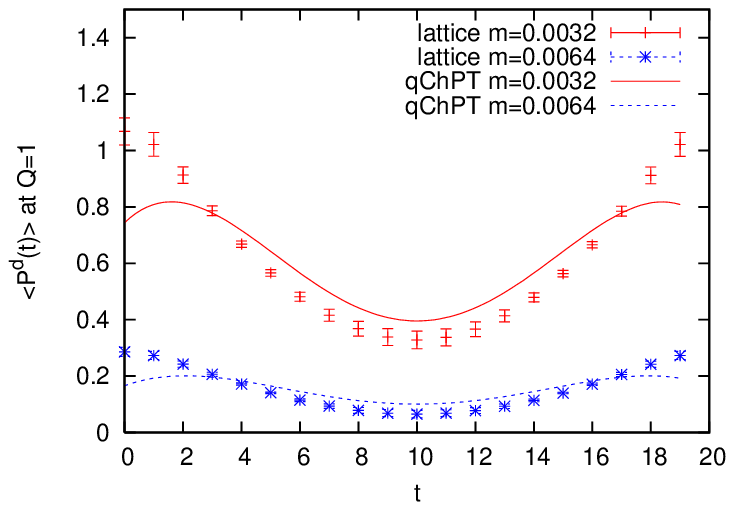}
  \,
  \includegraphics[width=65mm]{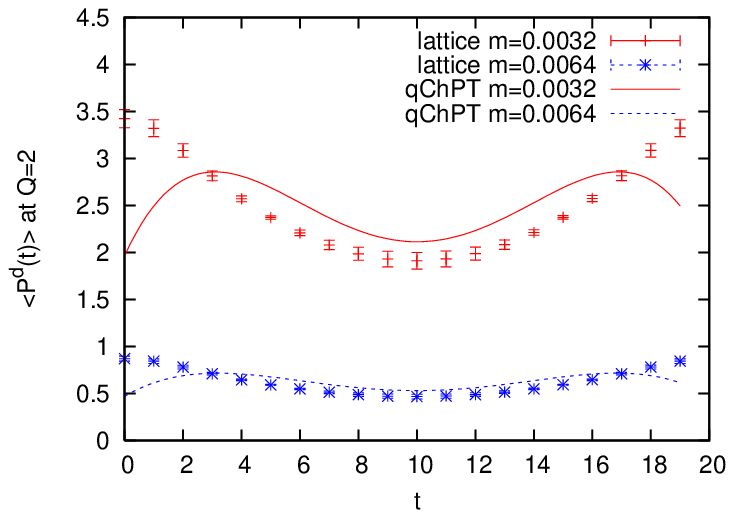}
  \caption{
    Disconnected pseudo-scalar correlators for
    |$Q|=1$(left) and $|Q|=2$(right). 
    The lines are the fitting function with the parameters
    which was obtained by calculating axial-vector
    current correlators and connected pseudo-scalar     and
    scalar correlators. 
  }
  \label{fig:disc}
\end{figure}

Data of the disconnected pseudo-scalar correlators are shown
in Figure~\ref{fig:disc}.
As in the connected correlator, we fit to the QChPT formula
with $\Sigma$ and $\alpha$ as free parameters.
The result for $|Q|$ = 0 and 1 data is 
$\Sigma^{1/3}$ = 227(32)(00)~MeV and 
$\alpha$ = $-$3.5(1.2)(0.3), which is consistent with
the connected correlators.
The analysis with $\alpha$ = 0 yields
$F_\pi$ = 125.7(5.6)(0.9)~MeV 
and $\Sigma^{1/3}$ = 223(29)(00)~MeV.

\section{Summary}
We performed a quenched simulation in the $\epsilon$-regime
at small quark masses $m$ = 2.6--13~MeV.
In this small quark mass region, the connected scalar and
pseudo-scaler correlators are approximated by the low-lying 
modes of the Dirac operator to 1\% accuracy or even better.
In the small topological sector, the results are well
described by the one-loop QChPT formula; by fitting we
can extract $F_\pi$ and $\Sigma$ as well as $m_0^2$ and $\alpha$.

We also find problems:
Data at higher topological sectors are not consistent with
the QChPT expectation.
Even for the small $|Q|$ sector, the large negative value of
$\alpha$ may suggest a breakdown of the theoretical
expression.
We suspect that these problems arises from the small
physical volume of our lattice for which the condition 
$|Q|\ll \langle Q^2\rangle$ is not well satisfied except for
$Q=0$.


\begin{thebibliography}{99}

\bibitem{Gasser:1987ah}
J.~Gasser and H.~Leutwyler,
\emph{Thermodynamics Of Chiral Symmetry}
Phys.\ Lett.\ B {\bf 188}, 477 (1987).

\bibitem{Damgaard:2001js}
P.~H.~Damgaard, M.~C.~Diamantini, P.~Hernandez and K.~Jansen,
\emph{Finite-size scaling of meson propagators}
Nucl.\ Phys.\ B {\bf 629}, 445 (2002)
[arXiv:hep-lat/0112016].

\bibitem{Damgaard:2002qe}
P.~H.~Damgaard, P.~Hernandez, K.~Jansen, M.~Laine and L.~Lellouch,
\emph{Finite-size scaling of vector and axial current correlators}
Nucl.\ Phys.\ B {\bf 656}, 226 (2003)
[arXiv:hep-lat/0211020].

\bibitem{Fukaya:2005yg}
  H.~Fukaya, S.~Hashimoto and K.~Ogawa,
  \emph{Low-lying mode contribution to the quenched meson correlators in the
  epsilon-regime}
  arXiv:hep-lat/0504018.

\bibitem{Ogawa:2005jn}
  K.~Ogawa and S.~Hashimoto,
  \emph{Effect of low-lying fermion modes in the epsilon-regime of QCD}
  arXiv:hep-lat/0505017.

\bibitem{Bietenholz:2003bj}
  W.~Bietenholz, T.~Chiarappa, K.~Jansen, K.~I.~Nagai and S.~Shcheredin,
  \emph{Axial correlation functions in the epsilon-regime: A numerical study  with overlap fermions}
  JHEP {\bf 0402}, 023 (2004)
  [arXiv:hep-lat/0311012].

\bibitem{Giusti:2004yp}
  L.~Giusti, P.~Hernandez, M.~Laine, P.~Weisz and H.~Wittig,
  \emph{Low-energy couplings of QCD from current correlators near the chiral limit}
  JHEP {\bf 0404}, 013 (2004)
  [arXiv:hep-lat/0402002].

\bibitem{Bietenholz:2005ip}
  W.~Bietenholz and S.~Shcheredin,
  \emph{Overlap hypercube fermions in QCD with light quarks}
  arXiv:hep-lat/0508016.

\bibitem{Shcheredin:2005ew}
  S.~Shcheredin and W.~Bietenholz,
  \emph{Low energy constants from the zero mode contribution to the pseudo-scalar
  correlator}
  arXiv:hep-lat/0508034.

\bibitem{DeGrand:2004qw}
  T.~DeGrand and S.~Schaefer,
  \emph{Improving meson two-point functions in lattice QCD}
  Comput.\ Phys.\ Commun.\  {\bf 159}, 185 (2004)
  [arXiv:hep-lat/0401011].

\bibitem{Bardeen:2003qz}
W.~A.~Bardeen, E.~Eichten and H.~Thacker,
\emph{Chiral Lagrangian parameters for scalar and pseudoscalar mesons}
Phys.\ Rev.\ D {\bf 69}, 054502 (2004)
[arXiv:hep-lat/0307023].








\end{thebibliography}
\end{document}